**Tracking plumbing system dynamics at the Campi Flegrei caldera, Italy: high-resolution trace element mapping of the Astroni crystal cargo**

Rebecca L. Astbury[1*], Maurizio Petrelli[1], Teresa Ubide[2], Michael J. Stock[3] Ilenia Arienzo[4], Massimo D'Antonio[4,5], Diego Perugini[1]

[1]**R. L. Astbury \*(corresponding author)** – Petro-Volcanology Research Group, Department of Physics and Geology, Università degli Studi di Perugia, Piazza Università, 1, 06100 Perugia, ITALY. rebeccalouise.astbury@studenti.unipg.it

[1]**M. Petrelli** – Petro-Volcanology Research Group, Department of Physics and Geology, Università degli Studi di Perugia, Piazza Università, 1, 06100 Perugia, ITALY

[2]**T. Ubide** – School of Earth and Environmental Sciences, The University of Queensland, Brisbane QLD 4072, AUSTRALIA

[3]**M. J. Stock** - Department of Earth Sciences, University of Cambridge, Downing Street, Cambridge CB2 3EQ, UNITED KINGDOM

[4]**I. Arienzo** – Istituto Nazionale di Geofisica e Vulcanologia, Sezione di Napoli Osservatorio Vesuviano, Via Diocleziano 328, 80124 Napoli, ITALY

[4,5]**M. D'Antonio** – Dipartimento di Scienze della Terra, dell'Ambiente e delle Risorse, University of Napoli Federico II, Via Vicinale Cupa Cintia 21, 80126 Napoli, ITALY

[1]**D. Perugini** – Petro-Volcanology Research Group, Department of Physics and Geology, Università degli Studi di Perugia, Piazza Università, 1, 06100 Perugia, ITALY

**Abstract**

The Campi Flegrei caldera (southern Italy) is one of the most hazardous volcanic systems on Earth, having produced >60 eruptions in the past 15 ka. The caldera remains active and its potential for future eruptions is high, posing a danger to the dense population living nearby. Despite this, our understanding of pre-eruptive processes and the architecture of the sub-volcanic system are poorly constrained. Here, we combine established petrological techniques, geothermobarometric evaluation, and high-resolution trace element crystal mapping, to present a multifaceted, coherent reconstruction of the complex pre-eruptive dynamics and eruption timescales of Astroni volcano located in the eastern sector of Campi Flegrei caldera. The Astroni volcano is an important case study for investigating plumbing system processes and dynamics at Campi Flegrei caldera because it produced the most recent (ca. 4 ka ago) Plinian eruption within the caldera (Astroni 6); current long-term forecasting studies postulate that a similar sized event in this location is a probable future scenario.

Geothermobarometric results indicate interaction between an evolved, shallow magma chamber, and a less evolved, deeper pocket of magma, in agreement with previous studies focused on the Astroni 6 eruption products. In addition, a range of textural and trace element zoning patterns point to a complex evolution of both magmas prior to their subsequent interaction. High-resolution trace element crystal maps reveal discrete zonations in compatible elements. These zonations, combined with knowledge of K-feldspar growth rates, highlight a recharge event in the shallow plumbing system a few hours to days before the Astroni 6 eruption.

## 1. Introduction

Unravelling the dynamic processes that operate within sub-volcanic plumbing systems is one of the primary challenges in petrology and volcanology. These pre-eruptive processes may include: (1) progressive crystallization of a parental magma (potentially) leading to the formation of crystal mushes, (2) magma chamber recharge events, (3) magma-host rock interactions, (4) ascent to the surface, and (5) eruption (Annen et al., 2015; Arienzo et al., 2016; Bachmann and Bergantz, 2008; Cashman and Giordano, 2014; Di Renzo et al., 2011; Druitt et al., 2012; Petrelli et al., 2018).

As magmas cool and crystallise, they precipitate mineral grains, which record crucial information on the architecture and the development of the volcanic plumbing system (Blundy and Cashman, 2008 and references therein). Exploring discrete textural and chemical variations (i.e. zonation) in these crystals can provide a wealth of information on pre-eruptive histories. Examples include, but are not limited to, constraints on pressure and temperature conditions, changes in volatile element content, and timescales of magma ascent to the Earth's surface. Crystal zonations can also archive information about open-system processes (e.g. refilling events and the mixing between magmas) and differentiation (Arienzo et al., 2010; Blundy and Cashman, 2008 and references therein; Calzolaio et al., 2010; Humphreys et al., 2006; Iovine et al., 2017; Masotta et al., 2013; Ubide et al., 2014; Ubide and Kamber, 2018). A comprehensive study of crystals from volcanic products can enable us to untangle the magmatic history and dynamic processes acting in the volcanic plumbing system before an eruption.

Astroni represents an excellent example of a volcano fed by a complex plumbing system in the Campi Flegrei caldera (CFc), southern Italy, one of the most hazardous active volcanic regions on Earth (Orsi et al., 2009). Although most activity at CFc has been characterised by single eruptive events, the Astroni volcano produced multiple eruptions closely spaced in both time and location (Isaia et al., 2004). Activity at the Astroni volcano occurred between 4345 and 4192 cal. yrs BP (Smith et al., 2011). Eruptive events were mainly trachytic to phonolitic in composition, dominated by explosive, mostly phreatomagmatic activity, with subordinate lava effusions. Variation of [11]B and radiogenic [87]Sr in products from Astroni 1 to 5 indicate mingling between two, isotopically distinct end-member magmas (Arienzo et al., 2015; Di Renzo et al., 2011; Tonarini et al., 2009). Products from Astroni 6 and 7 are more enriched in [11]B radiogenic [87]Sr, suggesting complete mixing between the same two end-members (Tonarini et al., 2009). Among the explosive events, the Astroni 6 eruption (A6 hereafter) was the largest in the eruption sequence, commencing with a Plinian phase resulting in a 20 km high eruptive column (Isaia et al., 2004).

Although it is commonly accepted that the plumbing system leading to the A6 eruption experienced magma mingling/mixing events (Arienzo et al., 2015; Di Renzo et al., 2011; Isaia et al., 2004; Tonarini et al., 2009), little is known about its dynamic evolution and the history of ascending magmas. In particular, a detailed reconstruction of the timescales of pre-eruptive processes is challenging and requires further investigation. This is fundamental to develop realistic models of the evolution of the volcanic feeding system. In this study, we aim to answer the following two questions: (1) what were the pre-eruptive dynamics of the A6 plumbing system? (2) What are the time constraints on the processes occurring in the feeding system before the A6 eruption? We combine high-resolution trace element mapping with thermometric and barometric estimates on crystals from the A6 eruption, to unravel the progressive evolution of the volcanic feeding system and track the ascent of magmas from storage regions to the Earth's surface. In contrast to major elements, trace elements are much less affected by changes in pressure, temperature and volatile content (Ubide and Kamber, 2018 and references therein), which makes them ideal for investigating the history of melts in volcanic systems. Our data shed new light on the architecture and evolution of the A6 plumbing system, as well as on the mechanisms and timescales leading to eruption.

## 2. Regional Setting

### 2.1. The Campi Flegrei caldera (CFc)

Located in the south of Italy, to the west of the Apennine mountain chain, the CFc (Figure 1a) is the principal landform of the Phlegrean Volcanic District (Orsi et al., 1996). The entire system spans around 13 km in width, with its southernmost portion submerged below the Gulf of Pozzuoli (Capuano et al., 2013). The present morphology in the CFc area is related to its two largest eruptions which produced the Campanian Ignimbrite (CI, ~39 ka; De Vivo et al., 2001) and the Neapolitan Yellow Tuff (NYT, ~15 ka; Deino et al., 2004). In the past 15 ka, volcanism within the CFc has been very intense with eruptive products ranging in composition from shoshonite to trachyte and phonolite (D'Antonio et al., 2007, 1999). Around 70 eruptions occurred in three epochs of volcanic activity, between ~15 and 10.6 ka (Epoch I), 9.6 and 9.1 ka (Epoch II), and 5.5 and 3.5 ka (Epoch III), all of which were generated by vents located either inside the NYT caldera or along its structural boundaries (Di Renzo et al., 2011; Smith et al., 2011). The most recent eruption at CFc was Monte Nuovo in 1538 AD (Di Vito et al., 2016 and references therein). The CFc is still widely active, as demonstrated by recent periods of bradyseismic activity from 1969 – 1972 and 1982 – 1984 inducing a 3.5 m uplift of the Pozzuoli Harbour (Orsi et al., 1999). This is further supported by increased hydrothermal activity and rising gas emissions from fumaroles, as well as ongoing resurgence (Chiodini et al., 2016; Del Gaudio et al., 2010).

### 2.2. The Astroni volcano

The Astroni volcano is located within the Agnano-Monte Spina (AMS) volcano-tectonic collapse zone (Figure 1a), which formed within the CFc during Epoch III, between 4345 and 4192 cal. yrs BP (Smith et al., 2011). Stratigraphically, Astroni deposits lie above the AMS tephra (4625 – 4482 cal. yrs BP; Smith et al., 2011) and directly below the Averno 2 Fossa Lupara tephra (4192 – 3978 cal. yrs BP; Smith et al., 2011). The preserved tuff ring of the Astroni volcano is the result of seven eruptions, of varying magnitude, over several decades (Isaia et al., 2004). Isaia et al. (2004) implemented a detailed study of the entire Astroni sequence, defining seven discrete eruptive units (1 to 7 from base to top) via their lithology, sedimentology, and petrology, as well as erosional unconformities and palaeosol units. Each unit is further divided into several subunits depending on clast size, petrology, and geochemistry. A6 itself is comprised of six subunits of mainly phonolitic composition (Figure 1b; hereafter referred to as subunits 6a-f, from base to top).

Long-term hazard assessment and eruption forecasting by Orsi et al. (2009, 2004) postulated that, in the wake of renewed activity at CFc, the style and magnitude of a medium-sized explosive eruption would be similar to that of A6. They also hypothesised that there is a high probability of eruption in the Agnano-San Vito area in the future (where the Astroni tuff cone is located; Figure 1a), and that it may be the beginning of "a series of events closely spaced in time" (Orsi et al., 2009, 2004).

## 3. Materials and Methods

### 3.1 Samples

This study focuses on subunits 6a, 6c, and 6f of the Astroni sequence (Figure 1b). Representative minerals and glasses were analysed on polished thin sections ($\sim$60 μm thick). Around 30 feldspar and $\sim$30 pyroxene crystal separates of particle size $\geq 0.5$ $\phi$ ($\phi$ = -$\log_2 D/D_0$; D is the diameter in millimetres) from pumice and scoria samples were also analysed for this investigation, the crystals were set in epoxy and polished in preparation for analysis.

### 3.2. Analytical Techniques

Textural features and compositional changes in the A6 crystal cargo were investigated using transmitted light microscopy and Back-Scattered Electron (BSE) imaging on a Zeiss LEO 1525 Scanning Electron Microscope (SEM), operated at 20 kV, at the University of Perugia.

Electron probe microanalysis (EPMA) was performed at the Department of Earth Sciences, University of Florence, on a JEOL JXA-8600 four-spectrometer instrument, using an accelerating voltage of 15 kV and a beam current of 10 nA. Pyroxene and feldspar crystals were analysed with a beam diameter of 10 μm, and a 15 μm beam was used for glass analysis. Additional EPMA analyses were performed at Leibniz Universität Hannover on a CAMECA SX-100 five-spectrometer instrument using the same operating conditions. From repeated measurements of well characterized reference materials (i.e., NMNH 117733 and NMNH 122142; Jarosewich et al., 1980), the precision for all major oxides was estimated at approximately 2%. Analyses were carried out at crystal cores and rims. Further analyses were conducted at 20 – 100 μm intervals between the core and rim depending on the size of the crystal, or where textural or compositional variations were evident (hereafter termed intermediate zones). Intermediate zones that displayed a more mafic composition than their neighbouring, chronologically older zones were later renamed as 'internal' or 'final recharge zones', depending on their location in a specific crystal.

Trace element single spot determinations and high-resolution trace element mapping were performed by laser ablation–inductively coupled plasma–mass spectrometry (LA-ICP-MS) at the Petro-Volcanology Research Group (PVRG) facility, Department of Physics and Geology (Perugia), using a Teledyne Photon Machine G2 laser system coupled to a Thermo iCAP-Q ICP-MS. Helium was used as carrier gas with Ar and $N_2$ added after the ablation cell to avoid plasma destabilization and enhance the instrumental sensitivity.

Single spot analyses were performed on feldspar and pyroxene core and rims, plus intermediate zones where complex compositional textures occurred. The analyses were performed at the exact coordinates of previously analysed EPMA spots. Glasses were analysed in the same fashion. Data were collected for Sc, Ti, V, Cr, Mn, Ni, Rb, Sr, Zr, Nb, Cs, Ba, Hf, plus Rare Earth Elements (REEs), using spot sizes of 20 or 30 μm, depending on the width of the zone being analysed, with a repetition rate and fluence of 8 Hz and 3.5 J/cm$^2$ respectively. Dwell time was 10 ms per analyte. The NIST 610 reference material, Si concentrations determined by EPMA, and USGS BCR2G glass were used as the calibration standard, internal standard, and quality control, respectively. Data Reduction was carried out using the Iolite v.3 software package (Paton et al., 2011). Under the reported analytical conditions, precision and accuracy are better than 10% (Petrelli et al., 2016, 2008)

Twenty-two representative feldspar and pyroxene crystals were imaged following the line rastering technique proposed by Ubide et al. (2015) to produce multi-elemental maps. A range of spot sizes (10 – 30 μm), scan speeds (4 – 12 μm/s) and fluences (3.5 – 4 J/cm$^2$) were implemented to find the best time/resolution trade-off. Repetition rate was fixed at 8 Hz. When creating crystal maps, the number of analytes was reduced to increase resolution, however some REE analytes were still included in the analysis to account for small inclusions within crystals, as well as the enveloping matrix glass. The specifications therefore vary for each mapped crystal and are noted in Section 4.3. Crystal maps and transects were produced using 'Images from selections' and CellSpace (Paul et al., 2012), respectively, and compositional zonation relationships were further interpreted using 'Images from selections' with Monocle (Petrus et al., 2017).

### 3.3. Geothermobarometry

Crystallisation pressure and temperature estimates were iteratively obtained using equations Palk2012 and Talk2012 (Masotta et al., 2013; Mollo and Masotta, 2014) for clinopyroxene in alkaline differentiated magmas. Whole rock compositions from Tonarini et al., (2009) were tested as potential equilibrium liquid pairs for clinopyroxene cores with a value of 4.5 wt.% $H_2O$ (Arienzo et al., 2016). Average measured matrix glass compositions from this study were tested as potential equilibrium liquid pairs for clinopyroxene rims, with a $H_2O$ content of 2 wt.% (Supplementary Data 2). Crystal-liquid pairs were taken to be in equilibrium and used for geothermobarometry where they gave the best values of equilibrium activity ($0 \leq \Delta DiHd \leq 0.02$; Mollo and Masotta, 2014). Peaks in temperature and pressure were identified using kernel density estimates after the method of Sheather and Jones (1991). Corresponding depths were calculated ($P = \rho g h$) assuming a value of 2.3 g/cm³ for CFc crustal density after Rosi and Sbrana (1987). Feldspar rim crystallisation temperatures were estimated using two-feldspar thermometry (Equation 27b; Putirka, 2008), using an input pressure value of 1 kbar (Supplementary Data 2).

### 3.4. Timescale estimates

Timescale estimates for this study were achieved using a similar method to Ubide and Kamber (2018). We measured the thickness of the most recently recorded zones enriched in compatible elements, which are interpreted as final recharge zones and, when present, we also measured the additional thickness of the outermost rim after recharge. Measurements were consistently performed in the maximum growth direction along the *c*-axis of the crystal. Considering potential offsets due to sectioning effects, timescale results should be regarded as maxima. To undertake the measurements, crystal maps were exported from Cellspace (Paul et al., 2012) and distances were measured using ImageJ® software. A representative example of this procedure is provided in Supplementary Figure 1. In addition, trace element maps informed re-examination of BSE images for further thickness measurements using greyscale and LA-ICP-MS spot analyses. We assumed K-feldspar growth rates of between 10⁻⁸ and 10⁻⁷ cm/s, as reported by Calzolaio et al. (2010) for trachytic melts at low degrees of undercooling ($\Delta T$ between 10 and 100 °C), and hence obtained a range of timescales for each individual crystal.

## 4. Results

### 4.1. Textural characterization and geochemical constraints

Pumice samples of A6 have a low crystallinity (5 – 10 vol.%) with a moderate to highly vesicular glassy groundmass. The mineral assemblage is composed of K-feldspar (300 – 3000 μm), plagioclase (120 – 1500 μm), and light and dark green clinopyroxene (100 – 1300 μm) in descending order of abundance, with minor iron oxide (<300 μm) and biotite (<300 μm). Sparse aggregates of clinopyroxene + plagioclase ± biotite ± iron oxide are also observed (Figure 2).

### 4.1.1. Melt compositions

When compared to the range of whole rock and glass compositions found at CFc from the past 15 ka, A6 glasses have a relatively narrow compositional range (Figure 3a; Le Maitre, 2002), ranging from phonolite to trachy-phonolite; 56.1 – 60.4 wt.% $SiO_2$, 11.1 – 14.2 wt.% total alkalis ($Na_2O + K_2O$).

The average trace element values of glasses from subunits 6a, 6c and 6f measured by LA-ICP-MS single spot analyses are normalised to primitive mantle values (McDonough and Sun, 1995) and reported in Figure 3b. The compositional range for the entire Astroni sequence (Smith et al., 2011; Tonarini et al., 2009), as well as the CFc suite from the past 15 ka (D'Antonio et al., 1999), is also illustrated in Figure 3b. Glasses from each of the investigated subunits show similar trends, with depletion in Ba and Ti, and a relative enrichment in Pb. Overall, subunit 6a shows a slightly more pronounced depletion in Ba and higher concentrations in HREEs. Subunit 6f has a lower concentration of LREEs relative to 6a and 6c.

*4.1.2. Mineral compositions*

Textural investigation using BSE imaging allowed for the characterisation of clinopyroxene crystals into two groups: normal-zoned and reverse-zoned, the latter only noted in subunit 6a (Figure 1b, 4; Supplementary Data 1 contains major element data for representative crystal populations and glasses). Plagioclase crystals can be divided into three groups: normal-zoned, sieve textured (normal-zoned) and oscillatory-zoned, in order of abundance (Figure 4). K-feldspar crystals are categorised into four distinct populations: reverse-zoned, normal-zoned, oscillatory-zoned, and normal-zoned with a plagioclase core, in order of abundance (Figure 4).

Clinopyroxenes have diopside to ferroan diopside compositions ($Wo_{45-56}En_{25-48}Fs_{5-26}$) according to the classification scheme of Morimoto et al. (1988). Inter-elemental plots using magnesium number as differentiation index [$Mg\# = 100 * Mg/(Mg + Fe_t)$, Figure 5a] show that major element compositions from cores, intermediate zones, recharge zones and rims overlap with each other. There are slight distinctions between subunits with 6a containing the most evolved clinopyroxene compositions and 6c containing the most primitive (high Mg# and Cr-rich; Figure 5a) clinopyroxenes. Normal-zoned crystals show a slightly more evolved composition than the reverse-zoned population (Supplementary Figure 2).

Plagioclase is classified as bytownite to labradorite ($An_{42-92}Ab_{7-48}Or_{1-8}$) in composition according to Marshall (1996), with the majority of rims defined as labradorite and the majority of cores defined as bytownite. Intermediate zonations are equally spread between both compositions. K-feldspars are defined as a relatively homogenous sanidine composition ($An_{2-7}Ab_{15-23}Or_{72-83}$; Marshall, 1996). Bivariate plots of anorthite (%An) versus wt.% $Al_2O_3$ and FeO (Figure 5b,c) illustrate that plagioclase crystals have less evolved cores and more evolved rims. The relatively low %An cores shown in Figure 5b are plagioclase cores found within K-feldspar crystals. Oscillatory-zoned rims tend to be slightly less evolved in composition (Figure 4, Supplementary Figure 2). K-feldspar compositions do not correlate with previously defined textures, but some rims show a higher %An values (Figure 5c).

### 4.2. Pressure and temperature constraints on the Astroni system

Figure 6 illustrates crystallisation pressure and temperature for clinopyroxene cores and rims measured in this study that returned an equilibrium match with our potential equilibrium liquids (intermediate zones were not considered for the purpose of this figure). The results show a polybaric crystallisation trajectory, with crystal cores and rims formed at different pressures. Despite the partial overlap of results, kernel density plots show that the most probable conditions of clinopyroxene core crystallisation are at temperatures of ~975 °C (Standard Estimated Error [SEE] = 18.2 °C) and pressures ~1.7 kbar (SEE = 1.15 kbar), equating to a depth of around 7 km. Cores of reverse zoned clinopyroxenes did not return feasible equilibrium matches, and therefore only normal zoned clinopyroxene cores are represented in Figure 6. The most probable pressure of clinopyroxene rim formation is ~0.8 kbar, equating to around 3.5 km depth. The probability distribution for rim temperatures is less well defined with peaks between 955 and 970 °C (Sheather and Jones, 1991).

Two-feldspar geothermometry (Equation 27b; Putirka, 2008) on feldspar rims gives temperatures ranging from 800 to 970 °C ± 30 °C, and the highest frequency of results gives temperatures of ~870 °C ± 30 °C (Supplementary Data 2).

### 4.3. Trace element mapping

High-resolution trace element maps were created for representative clinopyroxene, plagioclase and K-feldspar crystals from subunits 6a, c and f, covering all textural groups identified petrographically (Figure 4, 7, 8 and 9). Single spot analysis of a larger set of crystals within each subunit was also performed, supporting our trace element crystal mapping results; this data is included as Supplementary Data 3. Most of the features revealed in these maps are not evident from BSE imaging and are not clearly defined through major element analyses, making trace element maps a powerful tool to unveil complexities not otherwise captured by established petrological techniques.

### 4.3.1. Clinopyroxene

Normal-zoned clinopyroxenes (Figure 4) exhibit two different trace element zonation patterns. In the first type, crystals show little to no enrichment in compatible elements between core and rim (Figure 7a). In contrast, the second type, found in crystals from subunits 6c and 6f, contains an intermediate zone, relatively enriched in compatible elements Cr and Ni (Figure 7b).

Reverse-zoned clinopyroxenes, which are only found in subunit 6a (Figure 1b, 7c) display a resorbed core with a high concentration of compatible elements such as Cr (1500 ppm) and Ni (100 ppm), mantled by an intermediate zone with lower compatible element concentrations, and a final rim (~20 μm) returning to enriched compatible element concentrations (100 ppm and 80 ppm, Cr and Ni respectively).

The Cr- and Ni-enriched zones (10 – 100 ppm and 20 – 80 ppm, respectively) found at the rim of crystals from subunit 6a (Figure 2, 7b) and as intermediate zones close to the rim in crystals from subunits 6c and 6f (Figure 2, 7c) are of comparable width (between 22 and 46 μm) and contain similar REE concentrations (Supplementary Figure 3a).

### 4.3.2. Feldspars

Sparse rounded sieve texture plagioclase crystals (Figure 4) have a rim enriched in Sr and Ba relative to their core (Figure 8a). Trace element concentration variations in euhedral sieve texture crystals (Figure 8b), as well as other plagioclase populations, tend to follow textural changes noted in BSE imaging and major element analyses (Figure 8b).

K-feldspar trace element maps (Figure 8c-f) highlight several distinct features undetected in the major elements. Normal-zoned K-feldspars (including those with a plagioclase core; Figure 4) that show no distinct textural zonation in BSE images contain a relatively large, low-Ba rounded core (500 – 3000 ppm Ba). Normal cores sometimes contain multiple zonations, highlighted clearly in Ba concentration maps (Figure 8c), or have a relatively constant Ba and Sr concentration (Figure 8d). The cores are mantled by high-Sr (1000 – 3000 ppm) and -Ba zones (1000 – 7000 ppm) (Figure 8c,d).

Similar to normal-zoned K-feldspars, reverse-zoned crystals (Figure 4) contain a rounded core, with relatively low concentrations in Sr and Ba (250 – 1000 ppm and 300 – 2000 ppm, respectively). Cores can also contain two to three zones of Sr and Ba at relatively low concentrations. This feature is displayed in Figure 8e by Ba maps of varied maximum concentration. Rounded cores are mantled by several wider zones, enriched in compatible elements, with a range of concentrations (800 – 18000 ppm).

Trace element zonation in oscillatory-zoned K-feldspars correlate with features defined through BSE imaging (Figure 4, 8f). Similar to reverse-zoned K-feldspar crystals and oscillatory-zoned plagioclase crystals, the zonation is most pronounced in Ba. Crystals contain a rounded, lower Ba and Sr core, with zonations oscillating between 3000 and 6000 ppm. The crystal rims are characterised by a higher concentration of Ba (~10000 ppm). All of the mapped K-feldspars feature a final zone enriched in compatible elements at, or near to, their rim, which contains a consistent concentration of REEs (Supplementary Figure 3b). These zones are markedly wider in subunit 6f than in subunits 6a and 6c (Figure 8c-f).

**5. Discussion**

Previous geochemical studies of A6 examined petrological features (Isaia et al., 2004), major and trace element glass compositions (Perugini et al., 2015; Smith et al., 2011), and isotopic variations in bulk rock samples and crystal separates (Arienzo et al., 2015; Tonarini et al., 2009). Results pointed to a complex plumbing system where open-system mixing and mingling processes occurred between a pre-existing, more evolved magma, similar in composition to that of the NYT or AMS eruptions, and a less evolved, isotopically-distinct melt, which began erupting at CFc just prior to the onset of the Astroni sequence (Arienzo et al., 2015; Di Renzo et al., 2011; Isaia et al., 2004). Perugini et al. (2015) also investigated the diffusive fractionation process of trace elements, using end-member mixing experiments and numerical modelling on natural compositions. Results indicated very short mixing-to-eruption timescales, on the order of tens of minutes for A6.

### 5.1. Astroni 6 pre-eruptive dynamics

Although textural and major element analyses suggest that sub-volcanic systems undergo dynamic pre-eruptive processes, they provide only a limited picture and fail to fully demonstrate the complexity of these processes. Here, we combine these established techniques with state-of-the-art trace element crystal mapping and geothermobarometric estimates to resolve the pre-eruptive dynamics of A6. This approach reveals that the crystal cargo of A6 records at least one pre-eruptive recharge event and that the evolution of the system is polybaric, with eruptions sourcing magmas from different crustal storage depths (Figure 6-9).

Geothermobarometric evaluation of clinopyroxene cores and rims from Astroni 6 (Figure 6) illustrates that normal zoned clinopyroxene cores evolved at around 7 km depth (~1.7 kbar), corresponding to the main magma storage depths at CFc identified by seismic tomography (Zollo et al., 2008), and rims evolved at around 3.5 km (~0.8 kbar), equivalent to the depth of shallow magma bodies previously detected at CFc (see

Stock et al., 2018 and references therein). These results point to the recharge of a pre-existing evolved magma chamber at shallow crustal levels, by a less evolved magma ascending from deeper in the system, in agreement with previous studies on the Astroni sequence (Arienzo et al., 2015, 2010; Isaia et al., 2004; Stock et al., 2018; Tonarini et al., 2009). Normal zoned clinopyroxene crystal cores, with no change in trace elements between their core and rim (Figure 7a), grew in the main, deep storage region and were introduced into the shallower system during the recharge event, where they grew low pressure rims. Trace element compositions of A6 glass and major element concentrations of clinopyroxene suggest that the products erupted in subunit 6a are slightly more evolved than those from subunits 6c and 6f (Figure 3, 5a, 6). This suggests that the process of magma recharge created a zoned magma chamber that was sequentially tapped from evolved shallower levels downwards by the ensuing A6 eruption (Figure 9).

It is evident that both magmas experienced separate complex activity prior to their final interaction before the A6 eruption (Figure 9), which may include mixing with other magmas. Sieve cores in plagioclase crystals (Figure 4, 8a,b) highlight the complex evolution of the less evolved magmatic component, as the occurrence of this texture, alongside a change in trace elements (Figure 8a,b), can indicate interaction between the crystal's host magma and a more mafic melt composition (Humphreys et al., 2006; Nakamura and Shimakita, 1998; Streck, 2008). The sieve textures found in A6 plagioclases occur at the crystal core, therefore suggest that mixing between the less evolved magmatic component and an even more mafic magma would have occurred earlier in time, prior to its interaction with the evolved, shallow magma. With regards to the history of the evolved shallow magma, both clinopyroxene and K-feldspar crystals hint at complexity. Clinopyroxene crystals containing final recharge zones were evidently present in the shallow chamber prior to its interaction with the less evolved, ascending magma. However, reverse zoned clinopyroxenes with recharge rims contain resorbed Cr-rich cores (Figure 7c), which may have been sourced from depth earlier in time and intruded into the shallow chamber. Resorbed cores in K-feldspar (Figure 8c-f), mantled by one or more zones enriched in Sr and Ba (Figure 8c-f), suggest crystallisation under open-system conditions in the shallow storage system before the final recharge event. However, the question remains whether the recharge resulting in the final composition of products from A6 occurred because of a single injection of melt from the deep system into the shallow storage region just prior to eruption, or whether multiple pulses occurred prior to the individual phases that emplaced the products found in each of the studied subunits.

Despite of the complexity of zoning patterns, we observe consistency in terms of the last recorded recharge event. Interestingly, the thickness of the outermost rim overgrowing the recharge rims increases through the eruptive sequence. In principle, this suggests that a single recharge event occurred between a pre-existing, evolved magma and a less-evolved magma just prior to the A6 eruption, with no further recharge events as the eruption ensued. Evidence to support this hypothesis is the presence of a single zone enriched in compatible elements, interpreted as a recharge zone, found at the rim of K-feldspar and clinopyroxenes from subunit 6a (Figure 7a, 8c, f). This indicates that recharge was recorded just prior to their eruption. The recharge zone is also found in crystals from subunits 6c and 6f but is progressively further from the crystal rim (Figure 7b, 8d). The similarity between these recharge zones in clinopyroxene crystals from each subunit (Supplementary Figure 3) suggests that these zones are likely chronicling the same event; implying that one injection occurred just prior to the onset of the initial Plinian phase of A6. Clinopyroxenes and K-feldspars from subunits 6c and 6f show evidence of further growth after the injection was recorded (Figure 7b, 8e), suggesting that they spent further time in the plumbing system before being erupted (Figure 9). This is supported by their position higher up in the A6 stratigraphy (Figure 1b).

We also note that these recharge zones in K-feldspars have a large variability in trace element concentration and therefore, we consider the possibility that multiple pulses of injection occurred just before each phase of A6 eruption (Figure 8c-f). Towards the rim, Ba concentration of these zones varies between ∼2000 and ∼10,000 ppm, even within the same subunit, and could be due to different recharge events by magmas with a variety of compositions. However, crystals with a higher concentration of Sr and Ba may have been closer to the injection point, crystallising from a less hybridised magma (Bergantz et al., 2015). Arienzo et al. (2015) suggested that the less evolved magma rising from depth may be tephri-phonolite in composition. However, mixing between this magmatic composition and a more evolved melt resulted in the eruption of a phonolitic magma. Hence, using partition coefficients of Ba between K-feldspar and phonolitic melts ($^{Kfs/L}D_{Ba}$ = 0.36 – 9.36; Wörner et al., 1983) and concentrations of Ba in tephri-phonolitic glass found in the Astroni sequence (1200 ppm; Smith et al., 2011), we estimate the range of possible Ba concentrations in K-feldspar rims growing from such a recharge event to be between ∼400 and 11,000 ppm, in agreement with our range of recorded concentrations. There are a few reverse-zoned K-feldspar crystals hosting zonations with Ba concentrations well outside of our calculated values around 20,000 ppm (Figure 8e), which could indicate interaction with a more primitive magma. This hypothesis is further supported by the comparison of recharge zones in K-feldspars; final recharge zones have similar REE concentrations, whereas internal recharge zones with a much higher concentration of Ba, found in some reverse zoned K-feldspars (Figure 4, 8e) have slightly lower levels of REEs (Supplementary Figure 3b), indicating a more mafic source of magma. These internal recharge zones overgrow resorbed inner cores, and are not the final recorded recharge event in crystals from any of the investigated subunits, likely representing a previous recharge event. We therefore favour the interpretation of a single final recharge event, from a less evolved magma, into the shallow reservoir just before the eruption of A6.

*5.2. Time constraints on the Astroni 6 eruption*

We have illustrated that A6 magmas underwent a complicated history, followed by a single recharge event just before eruption. Based on thickness measurements of final recharge zones and outermost rims, and growth rates of K-feldspar from Calzolaio et al. (2010) at low rates of $\Delta T$, we calculated the timescale between recharge and eruption. K-feldspar rims are calculated to have crystallised at an average temperature of 870 ±30°C (Equation 27b; Putirka 2008), within error of the upper limit of K-feldspar liquidus in evolved alkaline melts (840 °C; see Arzilli et al., 2016 and references therein). Therefore, we can assume very low rates of $\Delta T$ in the A6 shallow system. Results yield very short pre-eruptive timescales between ~1 hour to 6 days (Figure 10; Supplementary Data 4). Subunit 6f consistently yields the longest timescales (between 4 hours and 6 days; Supplementary Data 4). These short timescales are supported by the presence and texture of biotite in the erupted products (Figure 2). Sparse biotite crystals from A6 show clear evidence of resorption, indicating that they were unstable in their host magma before eruption; experimental data from Fabbrizio and Carroll (2008) suggests that biotite is stable at >1.35 kbar in evolved alkaline magmas but would breakdown within a few days of injection into the shallow chamber. The preservation of biotite in erupted products therefore indicates that magmas from the deep source were stored in the shallow chamber for a maximum of a few days, as a longer residence would result in complete dissolution.

Given the complexity and potential danger of the CFc system, and in the interest of future eruption forecasting in the area, it is important to compare our recharge-to-eruption timescale estimates with others for both Astroni system and the wider CFc area (Figure 10c-f). Timescales of shallow reservoir processes have been investigated using a variety of chronometric techniques. A6 was previously investigated using diffusive fractionation of trace elements in glasses, with mixing-to-eruption results calculated on the scale of tens of minutes (Perugini et al., 2015), in relative agreement with the lower boundary of our recharge-to-eruption estimates. Diffusion chronometry performed on sanidine crystals from the AMS eruption, which preceded the onset of the Astroni sequence, provided longer timescales of 2 – 60 years (Iovine et al., 2017). The apparent discrepancy between these diffusion timescales and our results may be related to the fact that AMS is a much larger scale eruption (0.9 $km^3$ Dense Rock Equivalent [DRE]; Orsi et al., 2009) in comparison to A6 (0.1 $km^3$ DRE; Isaia et al., 2004). Recent studies suggest that there is a direct correlation between the pre-eruptive mixing time and the subsequent volume of erupted magma; longer timescales for larger eruptions (Reid, 2003). Crystallisation experiments on the 1538 AD eruption of Monte Nuovo (Arzilli et al., 2016), which produced a smaller volume of magma to A6 (<0.1 $km^3$ DRE; Rosi and Sbrana, 1987), together with estimates on ascent rate based on the preservation of biotite (Fabbrizio and Carroll, 2008), yielded timescales of ascent from shallow storage (~4 km) to the surface to be on the scale of hours to two days (Arzilli et al., 2016), in agreement with our magma transfer results.

Di Vito et al. (2016) calculated magma accumulation and crystal residence times in the CFc shallow system to be between 100 – 300 years. Suggesting that, although the shallow magma chamber was remobilised over very short timescales, it may have been stored at shallow levels for much longer. By combining this information with our own recharge-to-eruption time estimates, and timescales with constraints on processes within the deep system, we are able to determine the rates of magma movement and pre-eruptive processes for different magma storage regions at CFc over the past 15 ka (Figure 9, 10). Using Astroni 1 as an example, Stock et al. (2018, 2016) provide timescale estimates for volatile saturation-to-eruption in the deeper plumbing system (7.5 – 8.5 km) of between 10 and 1000 days, suggesting that late-stage volatile saturation is responsible for the initial ascent of magma from the deep system. Our recharge-to-eruption timescales correspond to the shallow system (≤3.5 km) and show that ascending magmas recharged shallow magma pockets, mixing with the resident melts on short pre-eruptive timescales of hours to days. These constraints are of critical importance for monitoring future activity at CFc: evidence of magma movement, i.e. seismic activity, at ≥7 km may point to the mobilisation of a less evolved magma. Subsequent signals of magma movement at ≤3.5 km could indicate an impending eruption from a shallow magma source (accumulated over hundreds of years) on very short timescales of hours to several days.

## 6. Concluding Remarks

Through the combined use of textural evidence, geochemical analysis of glasses and crystals, and development of high-resolution trace element mapping of the A6 crystal cargo, we have produced a detailed reconstruction of the pre-eruptive events in the build up to the A6 eruption. Our results reveal the complex evolution of two separate magma bodies over time; (1) an evolved shallow magma chamber, represented by resorbed and intricately zoned K-feldspar cores in erupted products, (2) a less evolved magma represented by sieve textured plagioclase, which ascended from depth (∼7 km). Mixing of magmas from the deep storage region into the shallow magma system occurred in a single recharge event between ∼1 hour – 6 days before the beginning of the A6 eruption, which can be tracked via recharge zones in clinopyroxene and K-feldspar crystals. This recharge event resulted in the formation of a zoned shallow (∼3.5 km) magma chamber, which was sequentially tapped from the top downwards over the course of the eruption. Constructing a complete history of magmatic processes before and during the A6 eruption provides new insights into sub-volcanic processes at CFc and is of crucial importance for monitoring future activity in the region, which may be similar to that of Astroni.

**Acknowledgements**

We thank E. Braschi and L. Fischer for assistance at the electron microprobe facilities at Università degli Studi di Firenze and Leibniz Universität Hannover, respectively. We acknowledge constructive reviews by Matthew Loocke and Aniko Batki, as well as editorial handling by Andrew Kerr, which helped us improve the manuscript. This project was carried out under the ERC-funded CHRONOS Project (Project ID: 612776; Funded under: FP7-IDEAS-ERC). The MIUR-DAAD Joint Mobility Project: (57262582) supported microprobe analysis at Leibniz Universität Hannover. T. Ubide acknowledges support from The University of Queensland (UQECR1717581) and the Australian Geoscience Council and Australian Academy of Science (34th IGC Travel Grant). M Stock acknowledges support from a Junior Research Fellowship at Christ's College, Cambridge.

**Figure Captions**

**Figure 1:** (a) Map of the Campi Flegrei caldera (CFc) and its surrounding area. The extent of the Campanian Ignimbrite and the Neapolitan Yellow Tuff calderas are indicated in red and blue dashed lines respectively. The map also indicates the locations of the Astroni crater within the Agnano-Monte Spina volcano-tectonic collapse zone. The red polygon indicates the area at highest risk of new vents opening in the future of CFc as postulated by Orsi et al. (2004). Bottom left inset is a map of Italy indicating the location of the CFc in southern Italy. (b) Stratigraphic log of the Astroni 6 eruption, adapted from Isaia et al. (2004) and Tonarini et al. (2009).

Figure 2: Back scattered electron (BSE) images of A6 products; clinopyroxene [cpx], plagioclase [pl], K-feldspar [kfs], biotite [bt], and crystal aggregate with bt + pl + cpx + iron oxide (Fe-ox), in vesiculated glassy groundmass. Scale bars indicate 100 µm.

**Figure 3:** (a) Total Alkali Silica diagram (Le Maitre et al. 2002) showing the matrix glass compositions of Astroni subunits 6a, 6c and 6f, within the context of compositions from the CFc (D'Antonio et al., 1999; Tonarini et al. 2009). The graph also includes whole rock (WR) and glass data from the entire Astroni stratigraphy (black dots) (Smith et al., 2011; Tonarini et al. 2009). (b) Multi-element diagram of glasses from A6 analysed by LA-ICP-MS spot analysis. Dark grey background indicates WR and glass trace element geochemistry of the entire Astroni sequence (Smith et al., 2011; Tonarini et al. 2009). Light grey background is the trace element geochemistry of CFc over the past 15 ka (D'Antonio et al., 1999). All concentrations are normalised to primitive mantle values (McDonough and Sun, 1995).

**Figure 4:** Table illustrating the populations of clinopyroxene, plagioclase and K-feldspar found in Astroni 6 products, defined through textural analysis using BSE imaging. Labels

show the average Mg-number (Mg#) for clinopyroxenes, average orthoclase (%Or) in K-feldspars, and range of anorthite (%An) for plagioclases.

**Figure 5:** Bivariate diagrams of (a) Mg-number versus $SiO_2$, $TiO_2$, $Al_2O_3$ and $Cr_2O_3$ for clinopyroxenes, (b) %An versus $Al_2O_3$ and FeO for plagioclase feldspars, and (c) %An versus Al2O3 and FeO for alkali feldspars from Astroni 6. Dashed lines mark the compositional range found in each investigated subunit; green = 6a, blue = 6c, red = 6f.

**Figure 6:** Clinopyroxene-melt geothermobaromatry results (Masotta et al., 2013; Mollo and Masotta, 2014) for clinopyroxene cores and rims from Astroni 6. The standard estimated error (SEE) for pressure and temperature is 1.15 kbar and 18.2 °C, respectively, as indicated by error bars. Kernel density estimates (KDE) are calculated for cores and rims along the temperature and pressure axes. The dashed lines show previously identified magma storage depths at CFc from geophysical and independent petrological constraints (see Stock et al., 2018 and references therein).

**Figure 7:** Representative high-resolution Cr and Ni maps of clinopyroxenes from subunits 6a, 6c and 6f. (a) normal zoned clinopyroxene with no changes in compatible element s between core and rim, 15 µm spot at 6 µm /s; (b) normal zoned clinopyroxene with compatible trace element variation, 30 µm spot at 10 µm/s. (c) reverse zoned clinopyroxene with compatible element enrichment at its core and rim. All maps were created using a fluence of 3.5 $J/cm^2$, and a repetition rate of 8 Hz. White dashed lines indicate the glass-crystal boundary. Black scale bar = 100 µm.

**Figure 8:** Representative high-resolution Sr and Ba maps of feldspar crystals from subunits 6a, 6c and 6f. (a) sieve zoned plagioclase with relative enrichment of compatible elements at the rim, 20 µm spot at 8 µm/s; (b) sieve zoned plagioclase with oscillatory rim, 12 μm spot at 5 μm/s; (c) normal zoned K-feldspar, 30 μm spot at 12 μm /s; (d) normal zoned K-feldspar, 20 μm spot at 8 μm/s; (e) reverse zoned K-feldspa;, this section displays two Ba maps at different maximum concentration in order to enhance the multiple zonations at the crystal core, 30 μm spot at 10 μm/s; (f) oscillatory zoned K-feldspar, 30 μm at 10 μm/s. Maps were created using a fluence between 3.5 and 4 J/cm[2], and a repetition rate of 8 Hz. White dashed lines indicate the glass-crystal or crystal-crystal boundary. Black scale bar = 200 μm.

**Figure 9:** Simplified cartoon of the Astroni 6 plumbing system. The depth of each magma chamber is cited using kernel density estimate (KDE) peaks from cpx-melt geothermobarometry (Masotta et al., 2013; Mollo and Masotta, 2014; Sheather and Jones, 1991). The size and shape of the plumbing system, and the size of the Astroni tuff cone are not to scale**.**

**Figure 10:** Simplified graph comparing timescales of pre-eruptive recharge in the shallow system of CFc (~3.5 km) from (a) this study, with published timescale assessments from (b) the larger Agnano-Monte Spina eruption using diffusion chronometry (Iovine et al., 2017); (c) the 1538 eruption of Monte Nuovo using crystalling experiments (Arzilli et al. 2016); (d) the Astroni 6 eruption using diffusive fractionation of trace elements in glasses (Perugini et al., 2015). Timescales of (e) total magma storage time in the shallow chamber (Di Vito et al., 2016) and (f) volatile recharge and pressurisation of the deep system (7.5 – 8.5 km; Stock et al., 2016, 2018), are also provided to give a full picture of the processes at work in the CFc system and the timescales on which they occur.

**Figure1**

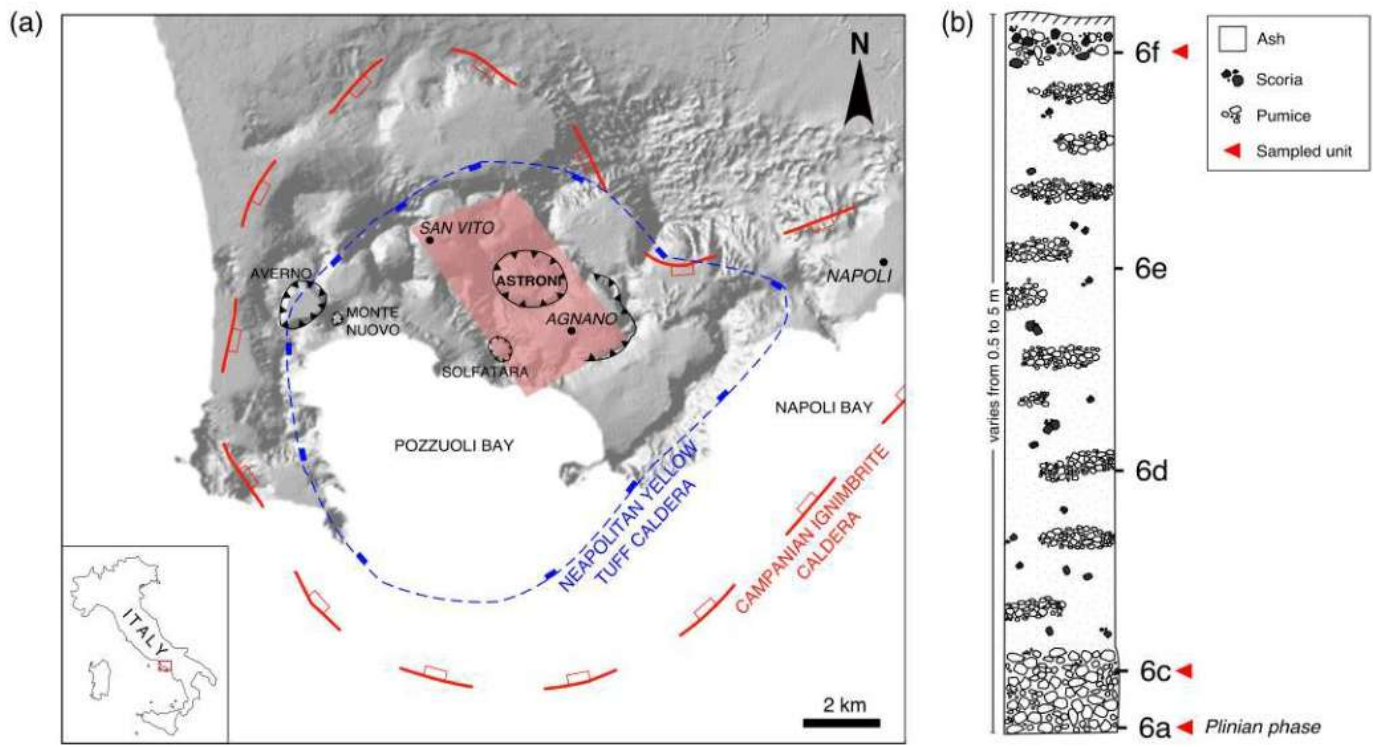

**Figure 1**

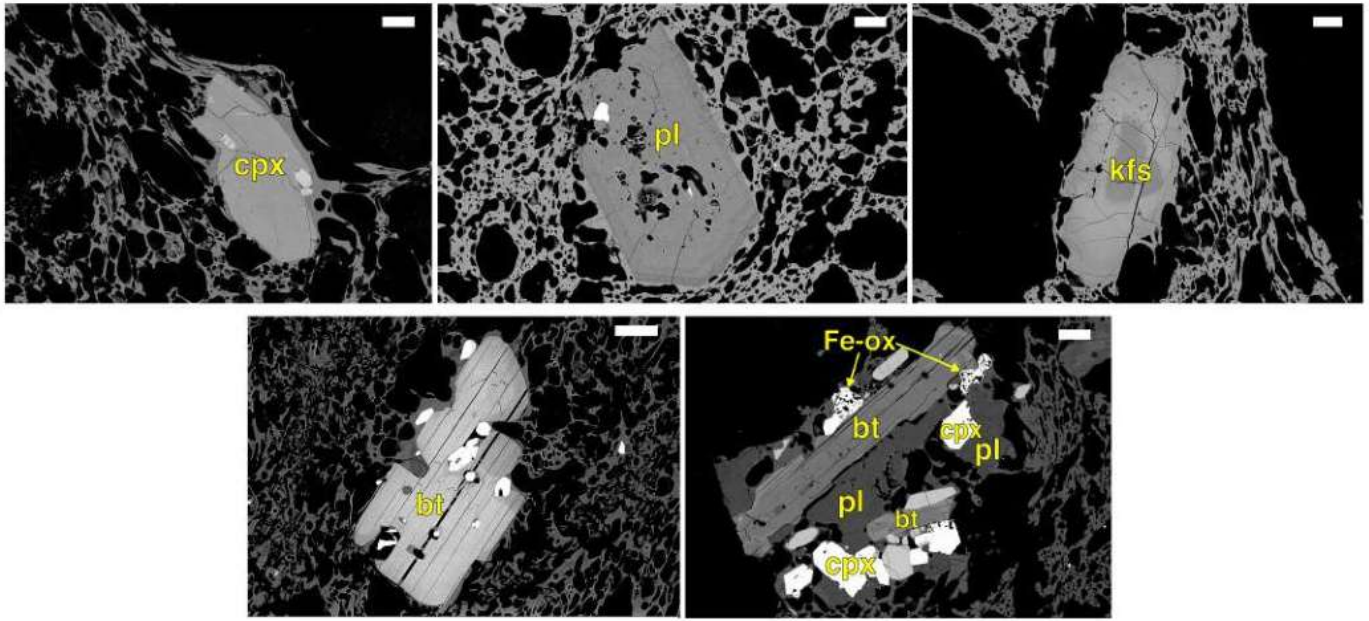

**Figure 2**

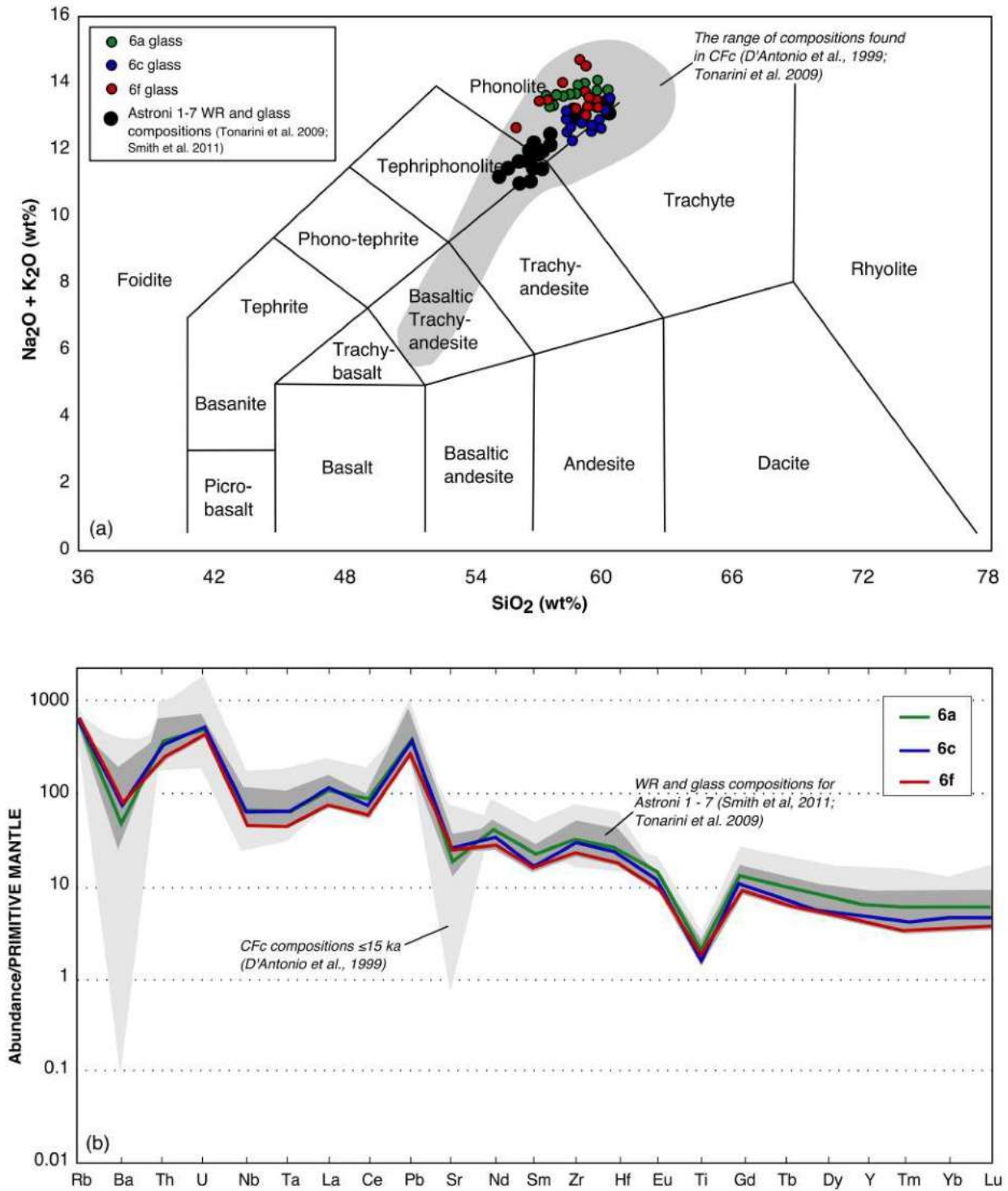

Figure 3

**Figure4**

| Mineral | Population Type | Representative Image | Description |
|---|---|---|---|
| Clinopyroxene | Normal | 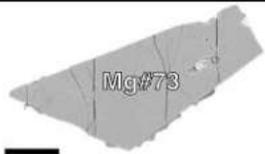 Mg#73 | Most dominant pyroxene population. Normal-zoned crystals maintain a consistent Mg-number between core and rim, some contain melt inclusions towards their core. Crystals vary in shape from euhedral to sub-rounded. |
| Clinopyroxene | Reverse | 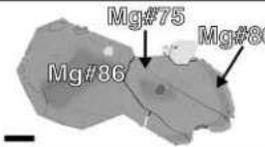 Mg#75 Mg#86 Mg#86 | Sparse population. Crystals consist of a high Mg-number core, mantled by a wider, more evolved zone. Rims then return to a high Mg-number composition. |
| Plagioclase | Normal | 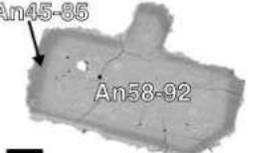 An45-85 An58-92 | Normal-zoned crystals generally contain a large, high anorthite core, with rounded edges. Cores are surrounded by a lower anorthite rim, with euhedral to sub-rounded edges. |
| Plagioclase | Sieve | 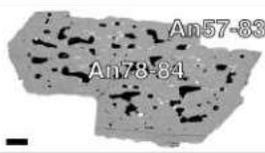 An57-83 An78-84 | Crystals have a wide, sieved core (moderate - high anorthite), mantled by a lower anorthite rim. Few crystals are completely intact, and vary in shape from euhedral to rounded. Many are found as intergrowths at the rim of large K-feldspar crystals. |
| Plagioclase | Oscillatory | 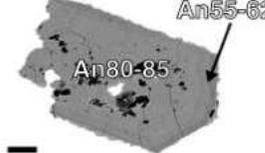 An55-62 An80-85 | Crystals consist of a wide high anorthite core, some have a sieve or dissolution texture. Core are mantled by lower anorthite rim, these rims consist of thin oscillatory zonations, and show some variation in anorthite. |
| K-Feldspar | Normal | 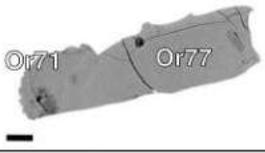 Or71 Or77 | Normal-zoned crystals show little to now change in texture or orthoclase content between core and rim. Most have sub-rounded to rounded edges. |
| K-Feldspar | Reverse | 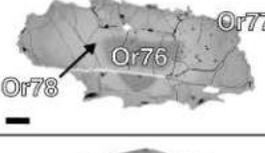 Or77 Or76 Or78 | Reverse-zoned crystals show little to no change in orthoclase between core and rim. In BSE they contain a darker, rounded core, mantled by a brighter zonation. This reflects Ba and Sr variations. |
| K-Feldspar | Oscillatory | 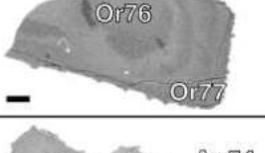 Or76 Or77 | Oscillatory-zoned crystals contain a wide, rounded, homogeneous core, mantled by several thin zonations towards the crystal rim. Orthoclase content remains constant across the crystal. |
| K-Feldspar | Normal with plagioclase core | 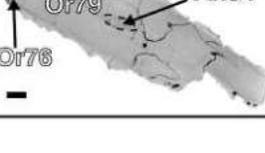 Or79 An54 Or76 | Normal-zoned crystals with a plagioclase core, tend to be elongated, and euhedral to sub-rounded in shape. Plagioclase core is small and rounded, with moderate anorthite content. Orthoclase content does not vary towards the rim. |

Figure 4

**Figure5**

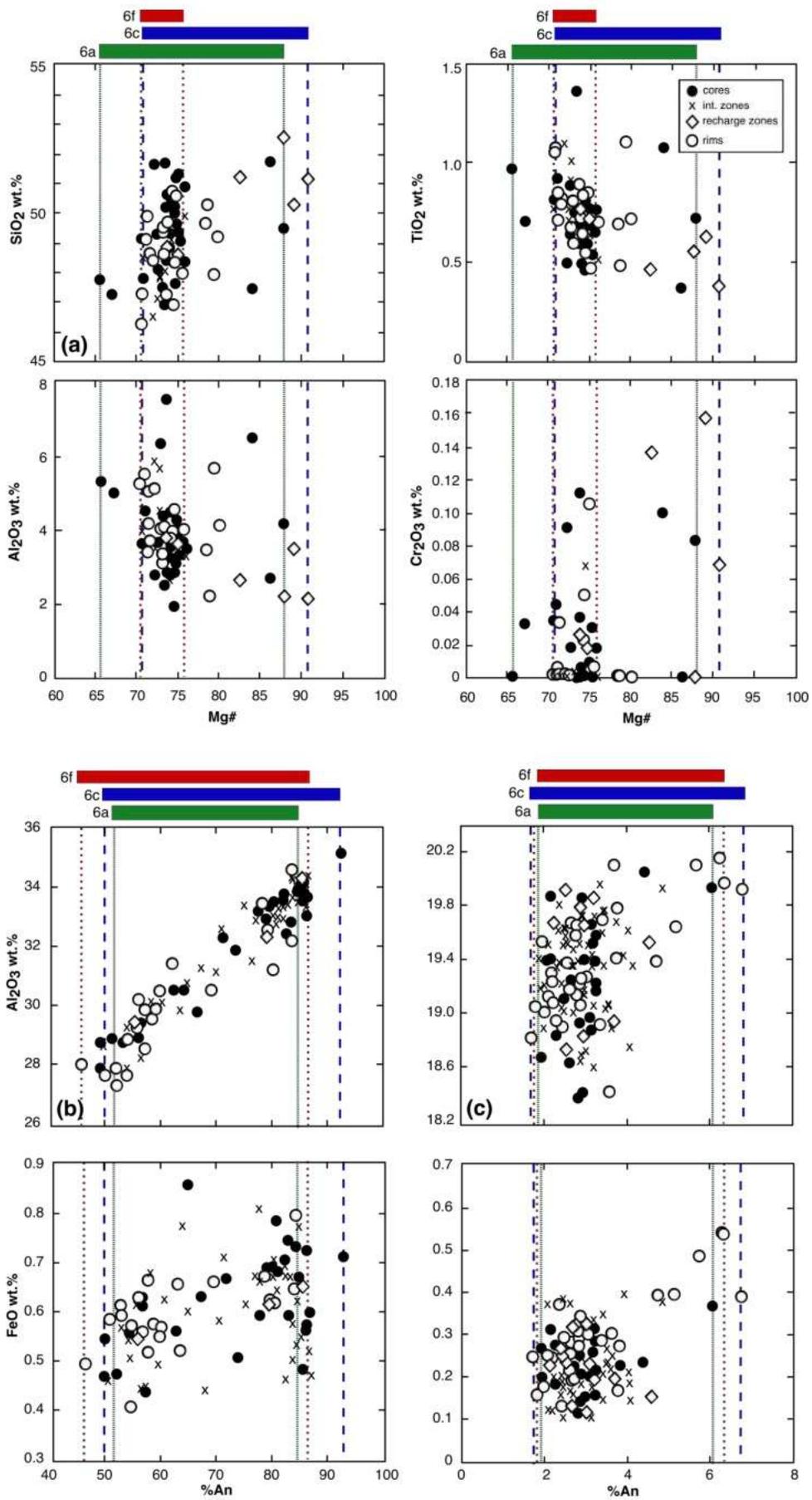

**Figure 5**

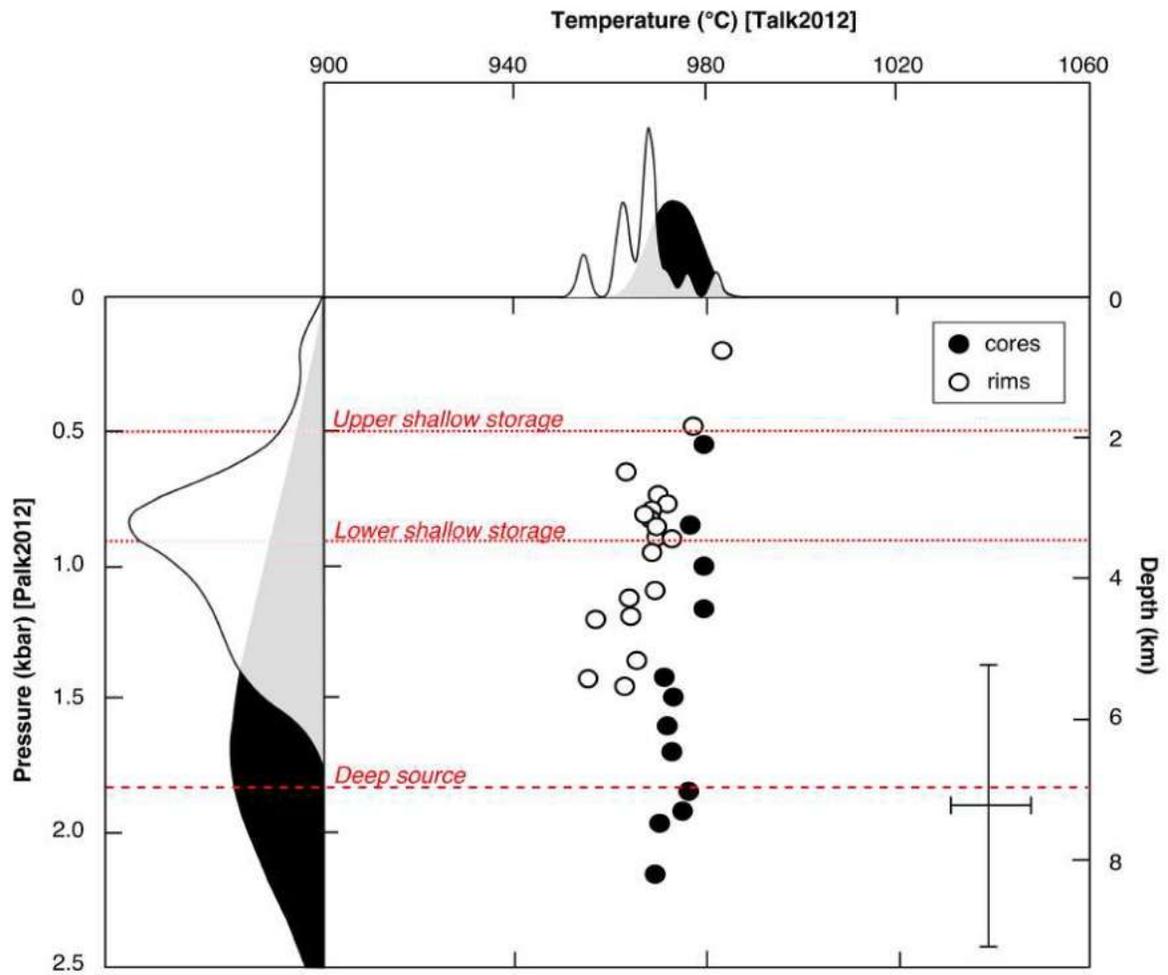

**Figure 6**

## (a) Normal zoned clinopyroxene
*6F_9*

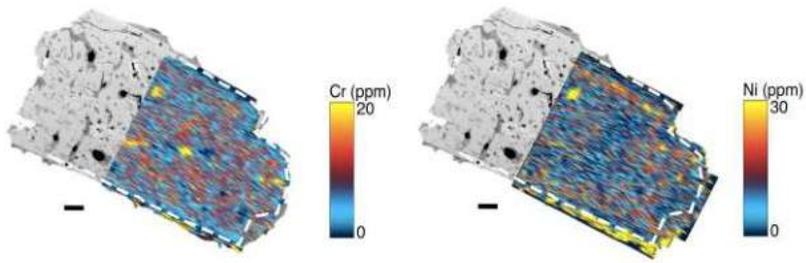

## (b) Normal zoned clinopyroxene
*6C_5*

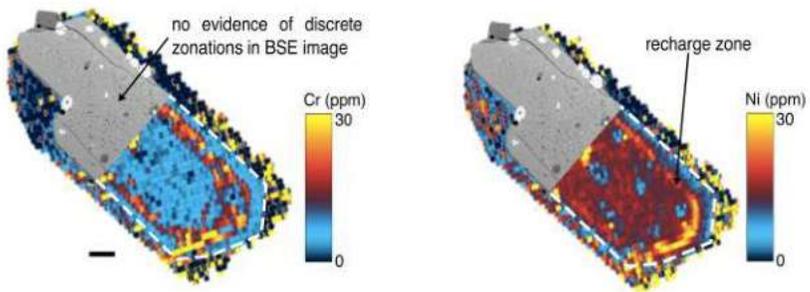

## (c) Reverse zoned clinopyroxene
*6A_13*

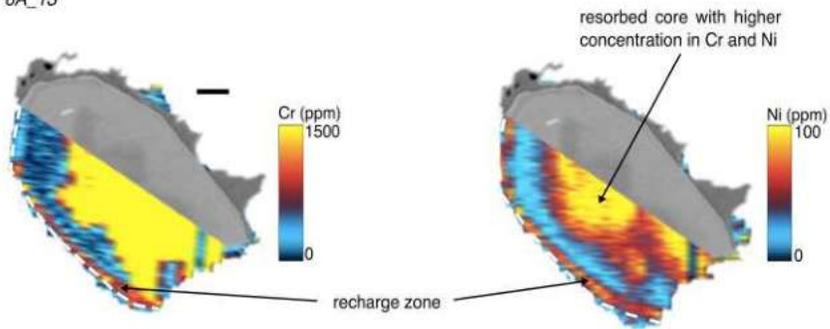

**Figure 7**

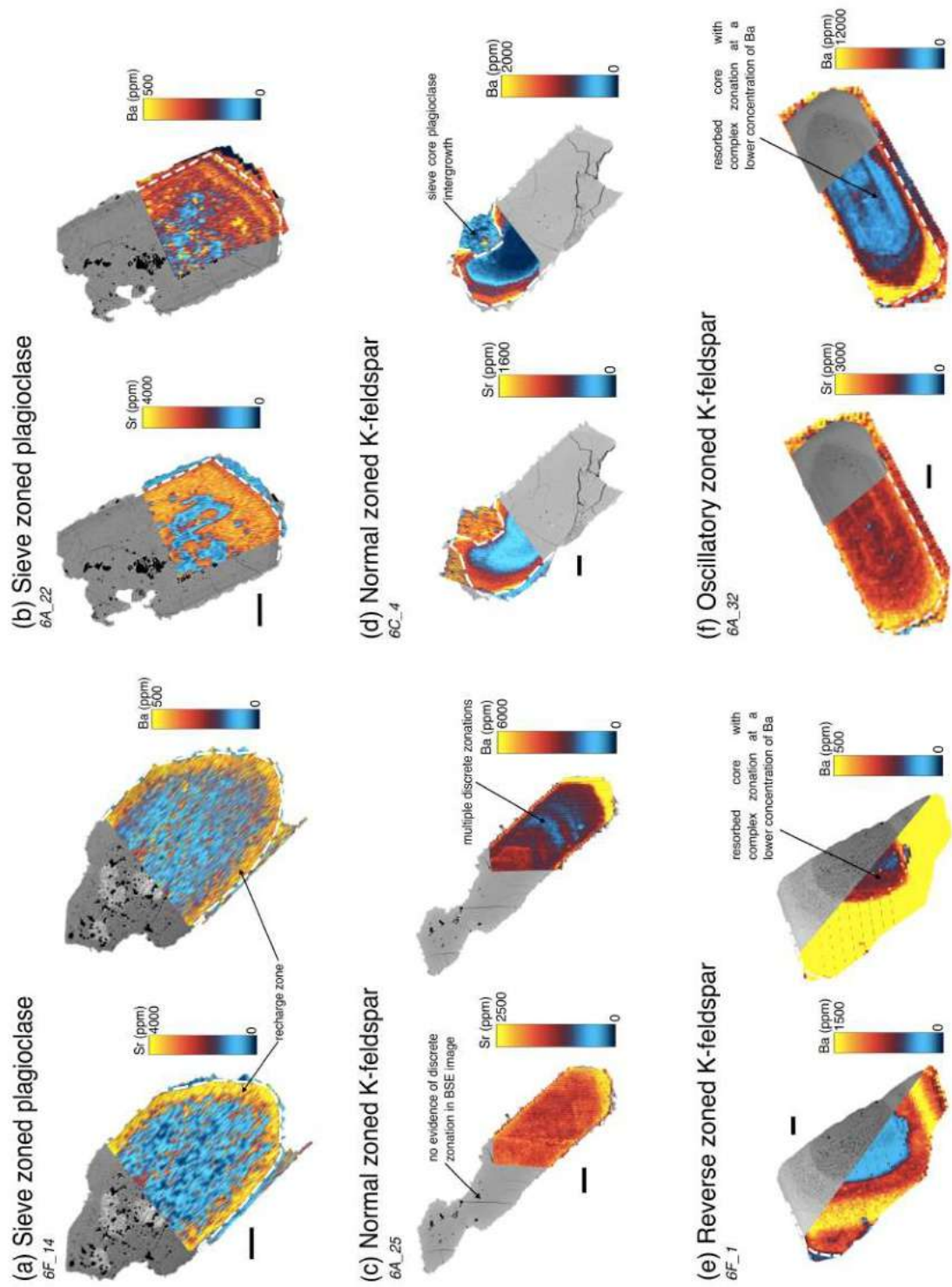

(a) Sieve zoned plagioclase
6F_14

(b) Sieve zoned plagioclase
6A_22

(c) Normal zoned K-feldspar
6A_25

(d) Normal zoned K-feldspar
6C_4

(e) Reverse zoned K-feldspar
6F_1

(f) Oscillatory zoned K-feldspar
6A_32

Figure 8

**Figure9**

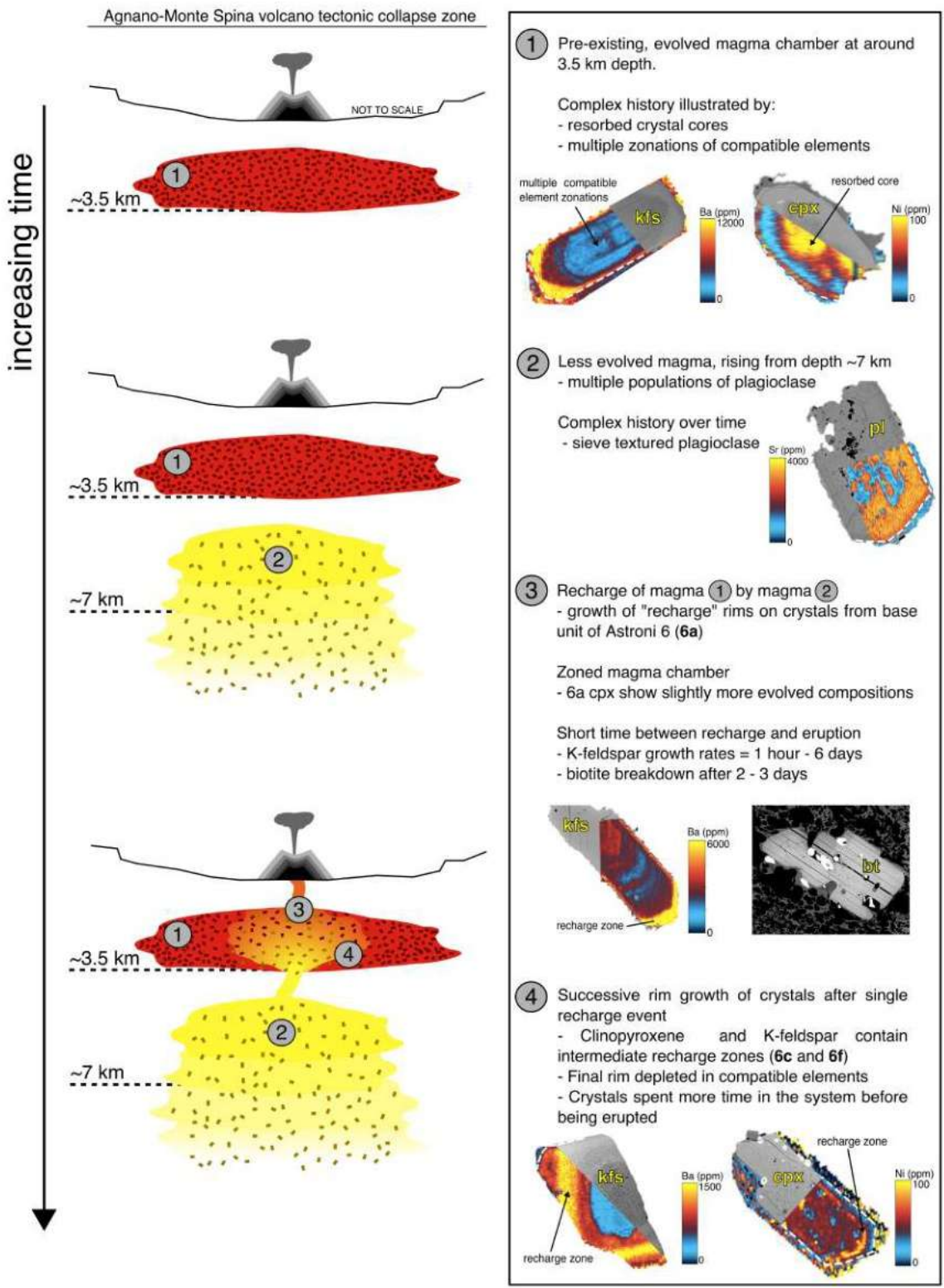

**Figure 9**

Figure10

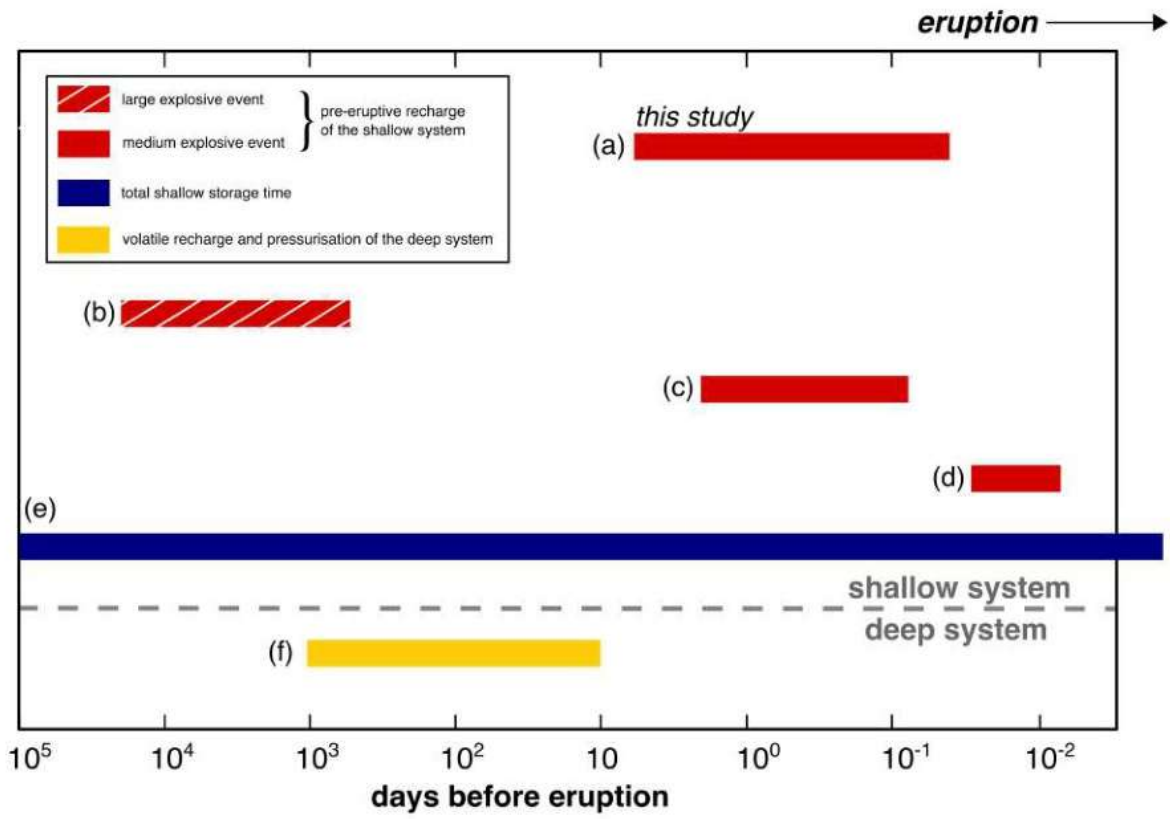

**Figure 10**

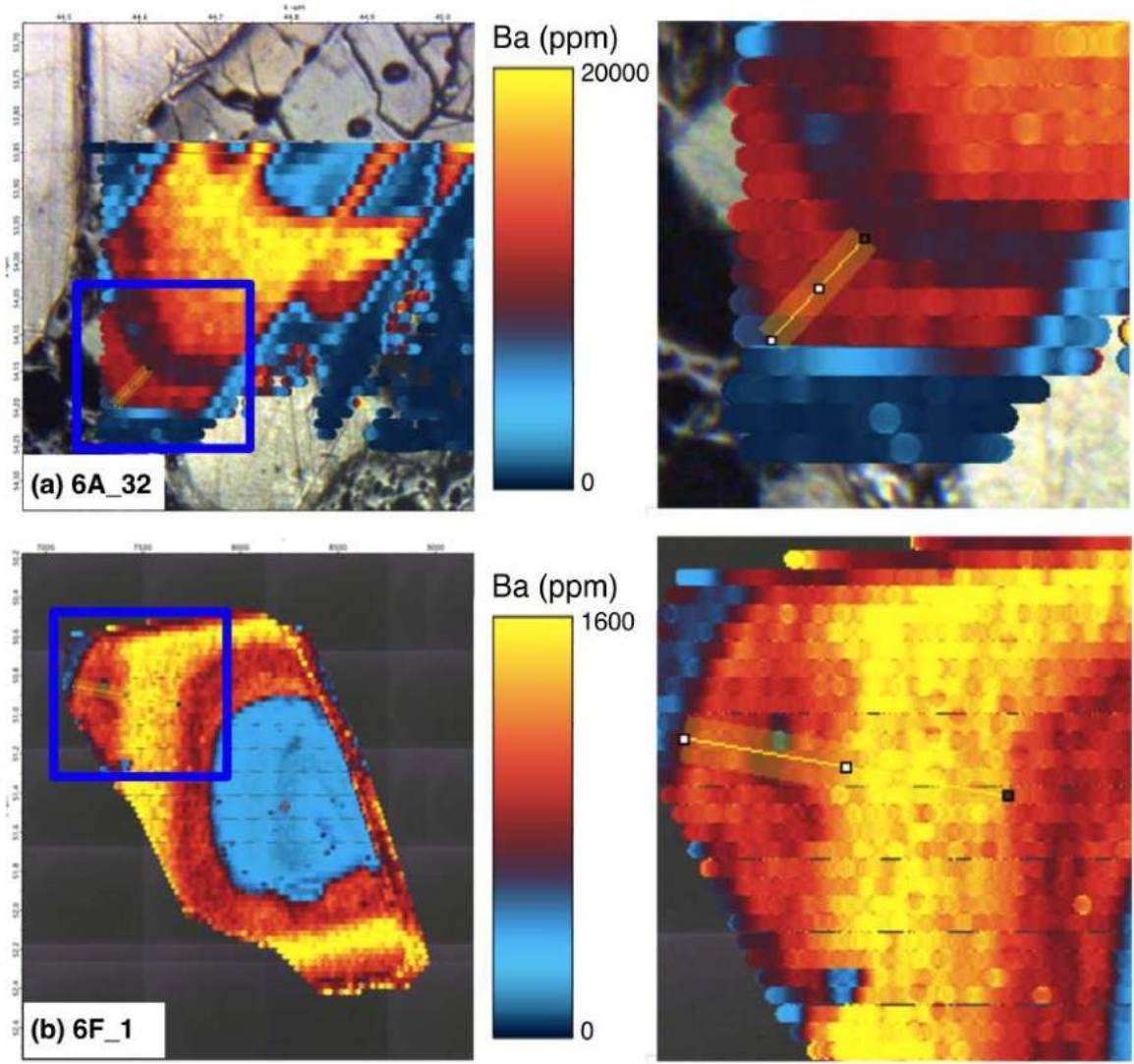

**Supplementary Figure 1:** Example image of measurement of maximum with of final recharge zone to edge of a K-feldspar crystal from subunits (a) 6a, where final recharge rims are found at, or extremely close to the rim, and (b) 6f, where final recharge zones are further from the crystal rim. Images were exported from Cellspace (Paul et al. 2012) and measured with ImageJ[®]

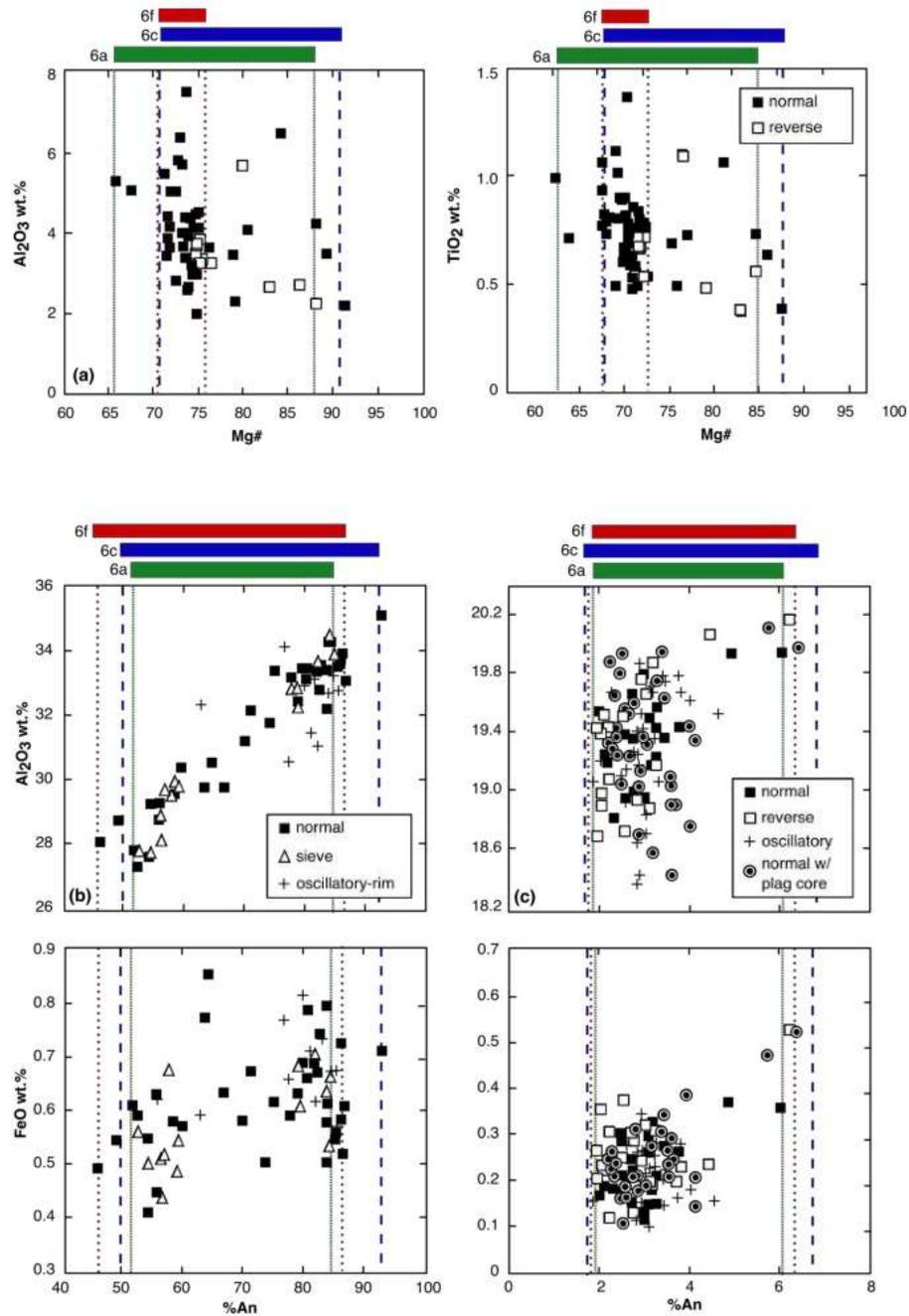

**Supplementary Figure 2:** Bivariate diagrams of (a) $Al_2O_3$ and $TiO_2$ versus Mg# for the different populations of clinopyroxene in Astroni 6, and $Al_2O_3$ and FeO versus anorthite content for the different populations of (b) plagioclase and (c) K-feldspar determined through textural analysis.

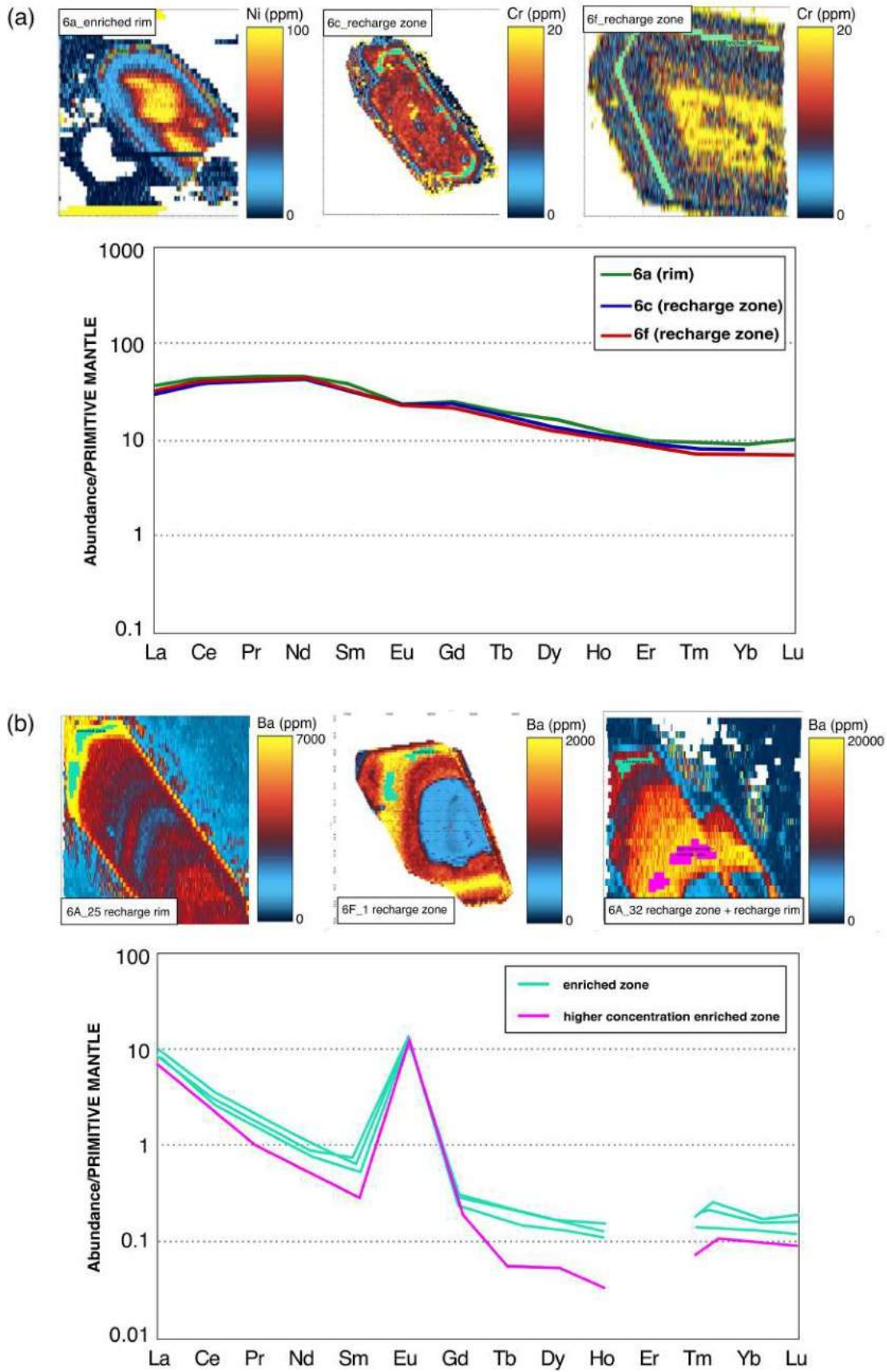

**Supplementary Figure 3:** Example images showing analysis of representative zones of enrichment in (a) clinopyroxene crystals and (b) K-feldspars from Astroni 6. Multi element diagram with values normalised to primitive mantle (McDonough and Sun, 1995) illustrates that the enrichment zones in (a) clinopyroxene are of similar REE concentration, and in (b) K-feldspars, final recharge zones are of similar REE concentration, but zones of a much higher concentration of Ba have comparatively lower REEs values.